# Sub-picosecond all-optical switching in a hybrid VO$_2$:silicon waveguide at 1550 nm

Kent A. Hallman[†,⊥,*], Kevin J. Miller[§], Andrey Baydin[†,⊥], Sharon M. Weiss[†,§,‡] and Richard F. Haglund[†,§]

[†]Department of Physics and Astronomy, Vanderbilt University, Nashville TN 37235 USA

[§]Interdisciplinary Graduate Program in Materials Science, Nashville, TN 37235 USA

[‡]Department of Electrical Engineering and Computer Science, Vanderbilt University, Nashville, TN 37235 USA

**Abstract**. Achieving ultrafast all-optical switching in a silicon waveguide geometry is a key milestone on the way to an integrated platform capable of handling the increasing demands for higher speed and higher capacity for information transfer. Given the weak electro-optic and thermo-optic effects in silicon, there has been intense interest in hybrid structures in which that switching could be accomplished by integrating another material into the waveguide, including the phase-changing material, vanadium dioxide (VO$_2$). It has long been known that the phase transition in VO$_2$ can be triggered by ultrafast laser pulses, and that pump-laser fluence is a critical parameter governing the recovery time of thin films irradiated by femtosecond laser pulses near 800 nm. However, thin-film experiments are not *a priori* reliable guides to using VO$_2$ for all-optical switching in on-chip silicon photonics because of the large changes in VO$_2$ optical constants in the telecommunications band, the requirement of low insertion loss, and the limits on switching energy permissible in integrated photonic systems. Here we report the first measurements to show that the reversible, ultrafast photo-induced phase transition in VO$_2$ can be harnessed to achieve sub-picosecond switching when small VO$_2$ volumes are integrated in a silicon waveguide as a modulating element. Switching energies above threshold are of order 600 fJ/switch. These results suggest that VO$_2$ can now be pursued as a strong candidate for all-optical switching with sub-picosecond on-off times.

Phase change materials, including vanadium dioxide (VO$_2$) and Ge$_2$Sb$_2$Te$_5$ (GST) glass, have recently attracted significant attention due to the large changes in their optical properties that can be triggered by external stimuli. These properties are being leveraged to achieve advances in on-chip components.[1,2] For volatile applications requiring ultrafast optical switching, VO$_2$ has long been considered one of the most promising candidates for integration with silicon photonics components. While experiments over the last decade have demonstrated that a VO$_2$ thin film can be optically switched on femtosecond time scales from the insulating to the metallic state and recover on a picosecond time scale,[3-5] most experiments were carried out with pump wavelengths near 800 nm, and none directly incorporated VO$_2$ with a silicon photonic component. However, ultrafast electron-diffraction studies comparing the photo-induced phase transition (PIPT) using near band-edge (2000 nm) excitation *vs* 800 nm pumping showed that the former requires only half the enthalpy change of the latter,[5] as seen in the observed monotonic decrease in PIPT threshold as the pump-laser photon energy approached the VO$_2$ band gap.[6] Hence, while prior work provides a rationale for integrating VO$_2$ into Si photonic components,[7-10] there is still no direct evidence that a VO$_2$:Si photonic element can operate on an ultrafast time scale when the optical pump and signal are in the telecommunications wavelength bands.

The incorporation of VO$_2$ into silicon waveguide devices is relatively recent. Switching at a modulation depth of 6.8 dB in a thermally switched ring-resonator device, with a VO$_2$ patch placed over a section of the ring, has been demonstrated.[7] Optical switching using a nanosecond Nd:YAG laser in an in-line waveguide and in ring resonators showed switching of a continuous laser signal at 1550 nm with moderately higher contrast ratios.[9] A more recent study of all-optical switching of 1550 nm femtosecond pulses in Si$_3$N$_4$ waveguide devices covered with a thin film of VO$_2$ showed switching times and contrast ratios to be proportional to waveguide length, with switching times of order 10 ps for a 10μm-long waveguide.[11]

Here we demonstrate sub-picosecond switching in an absorptive modulator comprising a silicon waveguide with an embedded section of VO$_2$, in which a 105 fs pulse at 1550 nm launched in a TE waveguide mode is switched by a temporally synchronized femtosecond pulse at 1670 nm, with an "on"-to-"off" time less than 1 ps. Figure 1(a) illustrates the essential concept of the experiment: a train of femtosecond pump pulses (red, 1670 nm) synchronously modulates the femtosecond probe (signal) beam (blue, 1550 nm). The results demonstrate that the optical dynamics in the confined geometry of the waveguide are more complex than what can be inferred from thin-film studies, but also indicate that the range of fluences traversed by the pump beam is consistent with a phase diagram[12] that shows an evolution from a transient metallic intermediate state that remains monoclinic to a phase in which the rutile crystal structure nucleates and grows, leading to longer recovery times. Thus, this experiment points the way to further developments that could lead to real progress in all-optical modulators with substantial advantages over purely electronic and electro-optic systems.

**DESCRIPTION OF EXPERIMENT**

Silicon waveguides with embedded VO$_2$ sections (green segment in Figure 1) were fabricated using silicon-on-insulator wafers (220 nm device layer, 3 μm buried oxide layer, SOITEC) using standard lithographic procedures, as reported in Ref [13]. Silicon waveguides with 700 nm gaps were defined by electron beam lithography (JEOL 9300FS-100kV) and subsequent reactive ion etching using a C$_4$F$_8$/SF$_6$/Ar gas mixture (Oxford Plasmalab 100). A second round of electron-beam lithography (Raith eLine) was used to open windows over the gaps in the waveguides for the VO$_2$ deposition. Then, VO$_x$ was deposited at room temperature by RF magnetron sputtering of vanadium metal at 6 mTorr total pressure with 20 sccm Ar and 1 sccm O$_2$.[14] After lift-off, the devices were annealed for 7 minutes at 450°C in 250 mTorr of O$_2$ to form polycrystalline VO$_2$ segments in the waveguide gaps. The lithography and VO$_2$ deposition processes were performed in two identical iterations to ensure complete O$_2$ diffusion during the anneal step. Figure 1(b) shows an example of one of the hybrid VO$_2$:Si waveguides. Atomic-force microscope (AFM) measurements on similarly prepared samples suggest that the average thickness of the VO$_2$ over the 700 nm-long waveguide gap is approximately 150 nm compared to the waveguide height of 220 nm.[13] All waveguides were cleaved to allow access for input and output butt-coupled tapered fibers (OZ Optics). The post-cleaved devices measured approximately 3 mm in overall length.

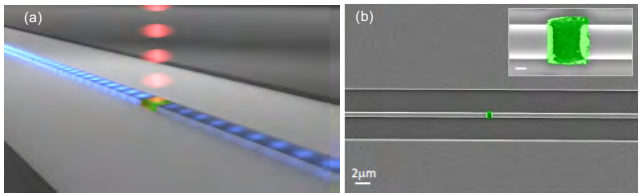

**Figure 1.** (a) Schematic view of the pump-probe experiment, showing 1550 nm femtosecond pulses (blue) injected into the waveguide from the right, and the gating femtosecond pulses (gold) at 1670 nm illuminating the embedded VO$_2$ segment from above. (b) False color SEM image of the embedded VO$_2$ modulator, 700 nm long, enlarged in the inset, where the scale bar is 200 nm.

Figure 2 shows the experimental configuration in which the dynamic response of the waveguide device was measured. An amplified titanium-sapphire laser system (Spectra Physics Spitfire Ace) producing nominal 100 fs pulses at 800 nm pumps an optical parametric amplifier (OPA, Light Conversion Topas) at 1 kHz repetition frequency. The OPA signal output is the probe beam that propagates through the waveguide; the idler output at 1670 nm is the free-space pump beam. A polarizing beam splitter separates the orthogonally polarized signal and idler beams at the OPA output. The temporal duration of the pump and probe pulses are derived from a spectral scan of, and a Gaussian fit to, the OPA output, as described in the Supplementary Information, Section S1. The characteristics of the pump and probe beams at the OPA output are summarized in Table 1.

The pump beam (red, Figure 2) is delayed with respect to the probe beam in a delay line comprising a corner cube on a computer-driven translation stage and focused onto the sample by a parabolic mirror. A pair of crossed polarizers on a computer-controlled rotation mount adjusts the fluence of the pump beam; to maintain the polarization of the pump beam fixed at the focusing mirror, only the upstream (laser side) polarizer is rotated, while the downstream polarizer remains fixed. The InGaAs detector for a pyroelectric power meter (not shown) on a flip mount in front of the parabolic mirror measures the average power of the pump beam, from which the energy per pulse and the fluence can be calculated.

The probe beam is chopped at 500 Hz, half the laser repetition rate, and this frequency provides the reference for phase-sensitive detection in the lock-in amplifier (LIA); the rationale for this configuration is discussed in the SI, Section S2. The input tapered fiber focuses the probe beam to a 2.5 μm spot and x-y-z piezoelectric actuators are adjusted to maximize the coupling of light from the tapered fiber into the proximal end facet of the silicon waveguide (220 nm × 700 nm). The average power coupled into the tapered fiber carrying the probe beam was estimated by exchanging the tapered fiber for an ordinary patch fiber connected to a fiber-coupled power meter. We estimate the insertion loss to be approximately 10 dB and confirmed this by measuring the transmission through a waveguide without embedded VO$_2$.

The embedded VO$_2$ segment, 700 nm long, is located near the middle of a silicon waveguide about 3 mm long, as shown by the scanning electron micrograph in Figure 1(b) where the embedded VO$_2$ segment is indicated in false-color green. The light transmitted through the VO$_2$ segment and onto the distal facet of the silicon waveguide is collected by the output tapered fiber, the alignment of which is controlled with similar piezoelectric actuators. Without further optical filtering or attenuation, the probe signal is then transmitted to a fiber-coupled indium gallium arsenide (InGaAs) detector (Thorlabs D400FC).

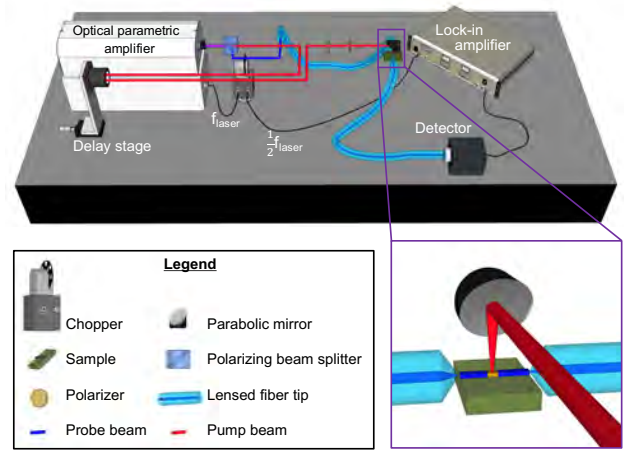

**Figure 2**. Experimental layout for the out-of-plane pump, in-waveguide probe configuration used to measure the temporal response of the embedded-VO$_2$ silicon waveguide (inset).

The pump beam is focused onto the waveguide by a parabolic mirror to avoid both chromatic and spherical aberration. The pump-beam and parabolic-mirror axes must be parallel to guarantee that translating the mirror does not distort the shape of the beam spot, and negligible differences between measurements when the parabolic mirror is swept parallel and perpendicular to the waveguide axis justify the claim that the axes are appropriately aligned. The waist of the roughly Gaussian pump-beam focus – 70 μm diameter at 1/e$^2$ of peak intensity – was measured by blocking the probe, changing the LIA reference frequency to 1000 Hz, and translating the parabolic mirror so the beam spot was swept across the device. In this configuration, scattered pump light coupled into the waveguide was measured by the LIA as a function of beam spot (parabolic mirror) position and fit to a Gaussian distribution. Given the dimensions of the waveguide, this implies that when perfectly aligned, it is reasonable to assume that the fluence is constant across the embedded $VO_2$. Information about the relevant optical properties of $VO_2$ are given in the SI, Section S3; relevant characteristics of the probe and pump beams are exhibited in Table 1.

**Table 1. Probe and pump-beam characteristics**

|  | Probe beam | Pump beam |
|---|---|---|
| Wavelength | 1550 nm | 1670 nm |
| Repetition rate | 500 Hz | 1000 Hz |
| Pulse duration | 105 fs | 130 fs |
| Pulse energy* | 0.5 pJ | 6 nJ – 360 nJ |
| Beam diameter | 220 × 700 $nm^2$ | 70 μm ($1/e^2$) |
| Fluence* | 32.5 $\mu J/cm^2$ | 0.13-7.8 $mJ/cm^2$ |
| Intensity | 33 $MW/cm^2$ | 1-60 $GW/cm^2$ |
| Insertion loss | 10 dB | 0 |

*At input face of waveguide

Figure 3 shows the differential transmission [$-\Delta T/T$ (%)] as a function of pump-probe delay, and illustrates the temporal switching behavior for the embedded-$VO_2$ silicon waveguides with a 700 nm long section of embedded $VO_2$. Figure 3(a), acquired at a moderate pump fluence of 2.7 $mJ/cm^2$, shows that the duration of the signal pulse reaching the InGaAs detector is no more than 640 fs FWHM. The pulse traces were fit by the sum of a Gaussian and a sigmoidal background.

Figure 3(b) shows the differential transmission for thirteen values of incident pump fluence, all acquired in a separate experiment and therefore probably under different alignment conditions than the measurement in 3(a). The photo-induced switching exhibits three distinct responses: (*i*) a low-fluence regime in which the signal returns to baseline in less than 1 ps (cool colors); (*ii*) an intermediate-fluence response in which there is rapid switching, but an elevated background after switching is complete (earth tones); and (*iii*) a high-fluence regime above roughly 5 $mJ/cm^2$ in which the peak amplitude continues to rise sublinearly but the background after switching also rises (warm colors).

The rising background as one progressively increases fluence through these three regimes shows that an increasing, fluence-dependent fraction of the $VO_2$ is in the rutile crystallographic state, in contrast to data at the low end of the fluence range, where the signal recovers quickly to the initial background level. The rising background in (*iii*) indicates the onset of the increasing fraction of the $VO_2$ section to the rutile crystallographic state and the long relaxation time associated with rutile-to-monoclinic relaxation, consistent with experiments at 800 nm.[3]

Figure 3(c) shows the differential transmission as a function of incident pump fluence at time $t$=0, incorporating the data from Figure 3(b) as well as other experiments. Details of the calculation of the absorbed energy density are discussed in the SI, Section S4. From previous studies,[3] approximately 10 ps after the pump pulse, either reversion to the semiconducting phase or the structural phase transition begins to take place, depending on the fluence.

Finally, we have calculated the full-width at half-maximum (FWHM) duration of the Gaussian component of the pump-probe dynamics for all data in Figures 3(a) and 3(b). Selected fits to data from Fig. 3(b), and for the FWHM as a function of fluence are shown in Figure 4. Computational details and fits to pump-probe spectra in Figures 3 and 4 are found in the SI, Sections S5 and S6.

For the brown square data points in Figure 3(d), extracted using the protocol in SI, Section 5, the FWHM signal rapidly decreases below 1 ps as the pump fluence is increased [Figs. 4(b), (c)], and is somewhat less than 1 ps at the highest fluences. However, the brown diamond points shown in that same figure, measured on a different day with what is presumed to be a different alignment, are substantially shorter, of order 640 fs – thus suggesting that the FWHM pulse overlap measured by the lock-in amplifier could be reduced well below 1 ps even at moderate pump fluences. The difference between the two sets of measurements could be a result of degradation of the $VO_2$ device during higher fluence irradiation, tuning of the OPA – and therefore pulse duration, or spatial chirp at the input and output silicon waveguide facets and its interaction with the lensed fiber tips.

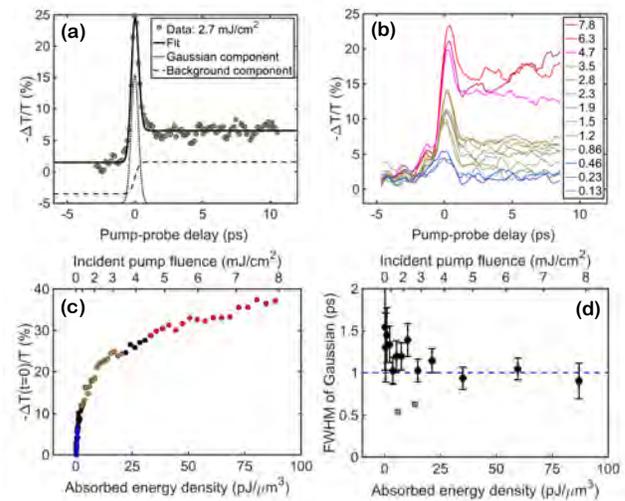

**Figure 3.** Differential transmission $-\Delta T/T$ (%) as a function of pump-probe delay (a) for an incident pump fluence of 2.7 $mJ/cm^2$ and (b) for a range of incident pump fluences as indicated by the legend on the right-hand side, (units of $mJ/cm^2$) showing the transition to the nearly fully switched rutile crystallographic phase at high fluences. The full width at half maximum of the pump-probe response in (a) is 0.64 ps assuming a Gaussian temporal profile. (c) Differential transmission *vs* incident pump fluence and absorbed energy density from below- to above-threshold levels for the monoclinic to rutile structural transformation, color-coded as in (b). (d) Duration of the Gaussian component of the detected signal from the pump-probe measurements in Figures 3(a) and (b), full-width at half maximum. The error bars represent the 95% confidence intervals from the fits to all of the unsmoothed data from Figure 3(b). The dashed blue line is at 1 ps. The much faster switching times represented by the brown diamond points were acquired at a different time than the black points [the data of Figure 3(b)], with adjusted alignment parameters; thus, they may represent the pulse duration at near-optimal alignment for the intermediate-fluence regime. Error bars for these two points are smaller than the plotting symbols. For clarity, the raw data for the curves in (b) were smoothed with a five-point moving average.

Figure 4 shows three pump-probe spectra selected from the measurements plotted in Figure 3(b) representing the modulator response at low, moderate and high pump fluences, as well as the pulse durations extracted for these spectra using the fitting procedure described in SI, Section 6.

At the lowest pump fluences [Fig. 4(a)], the embedded $VO_2$ is just barely switched. The pulse duration thus reflects the input probe pulse duration at the interface between waveguide and the embedded $VO_2$, and is dominated by the effects of dispersion in the input delivery fiber. Thus, the pulse duration reflected in the lock-in detector signal for Figure 4(a) can reasonably be considered to represent the instrumental response function. In Figure 4(b), in the moderate fluence regime, one sees the sharp onset of the gating pulse, but also the hint of a long-term response that does not recover to background within 10 ps. In Fig. 4(c), the overall trend of rising background with increasing pump fluence is fully developed; and high background relative to the peak signal indicates substantial metallization of the $VO_2$ modulator – foreshadowing nanosecond recovery.

Figure 4(d) shows the contrast ratio (%) in the differential transmission derived from the fits to the data shown in Figure 3(b), as a function of absorbed energy density in pJ/μm$^3$ and of *incident* pump fluence (the conventional measure of areal energy density in a pump-probe measurement). The contrast is defined as differential transmission at +8 ps subtracted from the peak differential transmission; both values were taken from the fit coefficients as described in more detail in the SI. The absorbed-energy scale is derived in section SI.4.

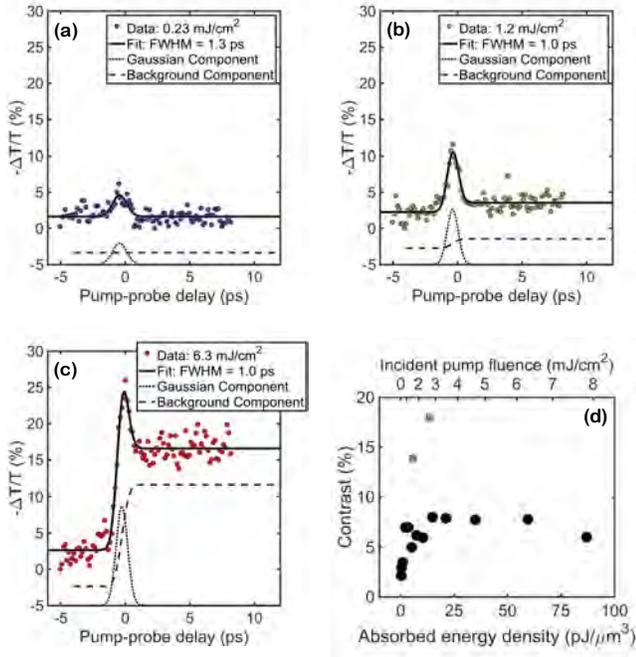

**Figure 4**. Differential transmission data (colored points) as a function of pump-probe delay, with fits to Gaussian pulse and background components for incident pump fluences of (a) 0.23, (b) 1.2 and (c) 6.3 mJ/cm$^2$. (d) Contrast ratio (%) in differential transmission as a function of absorbed energy density in pJ/μm$^3$ and of incident fluence in mJ/cm$^2$, derived from fits to data in Figure 3(b).

## DISCUSSION

The experiment described here builds on more than two decades of studies in which the ultrafast photo-induced phase transition (PIPT) in vanadium dioxide ($VO_2$) has been considered for potential application to all-optical switching technologies.[15-18] The PIPT in $VO_2$ generates both an insulator-to-metal transition (IMT) and a monoclinic-to-rutile structural phase transition (SPT). Evidence from many experiments on thin films of $VO_2$ [19-22] indicates that the IMT and SPT do not occur congruently, whether initiated ~~either~~ by local heating[23] or by nanosecond laser.[9, 24] Moreover, there is a critical threshold pump fluence above which relaxation to the monoclinic crystal structure takes nanoseconds. Finally, in a conventional pump-probe experiment, the pump and probe pulses have roughly the same duration, and the probe pulse is focused in a small fraction of the area illuminated by the pump to ensure that boundary effects do not distort the probe signal.

However, this experiment differs from conventional pump-probe studies in significant ways:

1. Assuming that the input tapered fiber has a dispersion of 15 ps/km·nm typical of fused silica, the 105 fs signal pulse from the OPA is stretched to a duration of about 1.2 ps FWHM at the waveguide input – and thus extends approximately 100 μm in space.
2. The transit time for any point on the phase front of the probe pulse through the 700 nm $VO_2$ modulator is of order 10 fs, negligible compared to the probe-pulse duration.
3. The absorption of the probe induced by the 135 fs pump pulse occurs during a small fraction of the probe-pulse duration as it traverses the embedded $VO_2$ segment.
4. The geometry of the $VO_2$ in this experiment is severely constrained compared to pump-probe studies on bulk thin films,[3, 20] with typical focal-spot volumes more than a thousand times greater than the volume of $VO_2$ that is embedded in the waveguide.
5. In this experiment, the sample is a device instead of a thin film. Therefore, volume absorbed energy density, as discussed in SI section 4, may be more relevant.

Against this background, we now propose the following global interpretation of the data in Figures 3 and 4. The linear absorption coefficient for monoclinic $VO_2$ over the spectral range spanned by pump and probe beams (SI, Section 1) is nearly constant, $\alpha_0 = 2$ μm$^{-1}$; absorption in the rutile phase is roughly an order of magnitude larger. The embedded $VO_2$ functions as an ultrafast absorptive modulator activated by the pump pulse. Because of the short pump-pulse duration, the absorption $-\Delta T/T$ rises rapidly to its peak; the initial slope of the pump-probe signal steepens with increasing fluence. The steepening is due to nonlinear absorption: even for pump fluences as low as 1 mJ/cm$^2$, the nonlinear absorption – at 800 nm equal to $\beta = 270 \pm 30$ cm/GW[26] – already exceeds $0.1\ \alpha_0$, so that this 10 *per cent* increase changes the absorption of the *probe* pulse in the embedded $VO_2$ by 20 *per cent*. Moreover, the probe pulse experiences this nearly instantaneous nonlinear absorptive effect throughout the entire 700 nm segment of $VO_2$.

With increasing fluence, the conduction-band electron concentration increases from $10^{20}$ cm$^{-3}$, and the fraction of unit cells excited in the embedded $VO_2$ begins to increase beyond the roughly 0.1 required for cooperative nucleation of the rutile phase;[6] significant metallization – and subsequent growth of the rutile phase – begins, leading to a rising background with increasing fluence, indicated by sigmoidal dashed curves that appear in the fits. The hot electrons and initially cold monoclinic lattice then equilibrate in a mixed rutile-monoclinic phase on a picosecond time scale.[21]

Measurements on $VO_2$ thin films excited by femtosecond pulses at 800 nm have shown that there is a fast-recovering metallic monoclinic transition that recovers on the time scale of a few picoseconds for incident fluences below 1-2 $mJ/cm^2$; as fluence rises from 1 to 5 $mJ/cm^2$, the post-switch background relaxes on an increasingly long time scale. This behavior also appears in thin films undergoing the ultrafast PIPT across a wide range of lattice mismatches and spatial correlation lengths.[3] This same behavior is evident in Figures 3(a), 3(b), 4(b) and 4(c): for incident fluences below 3 $mJ/cm^2$ the background signal returns close to baseline within the 10 ps measurement window, showing that at these fluences the $VO_2$ remains largely in the monoclinic phase[4, 19, 22] after the transient metallization phase.

The implications of these results for the present experiment are that (1) at an incident fluences of 1-3 $mJ/cm^2$, the embedded $VO_2$ has been only partially transformed from the monoclinic to the rutile crystalline phase, slightly slowing relaxation to the initial state; (2) the maximum value of contrast ratio between "on" and "off" states of the $VO_2$ will be reached near this threshold, (Fig. 3d); and (3) because this threshold electron density is reached by the end of the 135 fs pump pulse, we expect that the maximum probe-pulse signal also reaches the detector not long after the pump pulse is over, because at that point the absorbing rutile phase is well formed.

Moreover, the $VO_2$ section is thermally isolated by the $SiO_2$ substrate, and thus sheds the absorbed pump energy only very slowly by diffusive transport to the silicon waveguide. Given the confined geometry of the waveguide and the interface with the embedded $VO_2$, boundary effects on the waveguide and the substrate probably constrain the recovery time for the PIPT to the nanosecond time scale. Nonlinear pump-pulse effects on the silicon waveguide can be ruled out: Given the measured two-photon absorption $\beta_T \sim 0.5$ cm/GW and nonlinear index $n_2 \sim 2.8 \cdot 10^{-5}$ $cm^2$/GW for silicon at 1670 nm,[27] the fractional change in absorptive and refractive properties of silicon even at the highest intensities in this experiment is of order $10^{-4}$. This also rules out free-carrier absorption effects; the small change in $-\Delta T/T$ signal at the lowest fluence [Figure 4(a)] is due solely to the low density of electronic excitation.

Perhaps surprisingly, the dynamics for the 1670 nm pump-1550 nm probe sequence resemble those observed in optical pump (800 nm)-(white-light) probe experiments; in the latter, with pump-photon energies twice the bandgap energy of $VO_2$, one expects to generate a substantial population of hot phonons during relaxation down the ladder of excited states. In fact, however, the observed kinetics are also reasonably consistent with the pump-wavelength dependence at 2000 nm reported by Tao et al. Ref. [5]: in the high-fluence regime here – where transformation to the rutile phase is nearly complete – our measured values for the switching energy are of order 0.6 $eV/nm^3$, within a factor two of the value 1 $eV/nm^3$ measured in that ultrafast electron-diffraction experiment.

Therefore, the non-equilibrium dynamics of the photo-induced phase transition in even this simple in-line modulator structure poses crucial questions that remain to be resolved. For example, careful materials design could reduce relaxation times: substantially (200 times) shorter relaxation times from the photo-induced rutile phase have been measured for $VO_2$ thin films on MgO and $Al_2O_3$ compared to $SiO_2$ substrates for incident fluence near 8 $mJ/cm^2$.[28] In the in-line modulator, substrate effects at intermediate pump fluences should be explored, for example, by employing silicon-on-sapphire (SOS) substrates, since the thermal diffusivity of sapphire exceeds that of silica by more than an order of magnitude.

On the other hand, the phase diagram proposed by Cocker et al. from optical pump-THz probe measurements on thin $VO_2$ films on a-cut sapphire substrates suggests that the base substrate temperature might be even more important. Consistent with the present results, that phase diagram shows that at room temperature and at incident fluences of order 1.5 $mJ/cm^2$, the crystallographic phase transition nucleates, converting a fraction of the illuminated volume into the metallic phase with its long relaxation time; at fluences of 8 $mJ/cm^2$, in fact, the entire pumped volume of the film transitions immediately transitions into the rutile phase.[12] But at 225 K substrate temperature, picosecond relaxation to background is observed even for fluences approaching 4 $mJ/cm^2$. That temperature range could be reached by thermoelectric cooling – thus enabling the optimal ultrafast dynamics at higher fluence, albeit at the cost of modest additional power consumption.

## CONCLUSIONS

The results presented in Figures 3 and 4 demonstrate sub-picosecond switching of the in-line absorptive modulator with a maximum contrast of about 18%. Moreover, the graph of contrast ratio *vs* absorbed pump-pulse energy hints at a stable operating point over a modest range of low pump fluences. Thus, the switching performance described here, combining high intensity and low fluence and absorbed energy density, at laser wavelengths in the telecommunications band, approaches a reasonable operating range for ultrafast silicon photonics. However, the rising post-switch background beyond 10 ps for the higher fluences observed in this experiment would clearly be problematic for high modulator frequencies, because above the intermediate fluence range, so much of the embedded $VO_2$ transforms to the rutile crystalline phase that recovery to the initial state takes nanoseconds or longer – thus posing a critical obstacle to switching faster than a few GHz. This residual signal is almost certainly due to the pump beam, as the probe fluence – 33 $\mu J/cm^2$ – cannot initiate substantial metallization of the $VO_2$.

The dispersion lengthened duration of the input pulse, which is injected by a tapered silica fiber, does not appear to be a fundamental limiting factor in the performance of the in-line modulator. Thus, in the case of the in-line modulator, the pursuit of substantial reductions in switching time is now less urgent than optimizing the size, shape and volume of the $VO_2$ structure, as well as the thermal characteristics of the chip at high pump rates. Development of on-chip, in-plane optical pumping structures and investigation of alternate device geometries such as the ring resonator, pose additional significant and interesting challenges for the future.

## AUTHOR INFORMATION

**Corresponding Author**

*E-mail: KHallman@aegistg.com.

**Present Addresses**

⊥Kent Hallman: AEgis Technologies Group, 410 Jan Davis Drive, Huntsville, AL 35803 USA

⊥Andrey Baydin: Department of Electrical and Computer Engineering, Rice University, Houston, TX 70005 USA

**Funding Sources**

This research was supported in part by the National Science Foundation under grant EECS 1509740.

**Acknowledgements**

We thank Dr. Sergey Avanesyan (Fisk University) for help in resolving issues relating to the pump-probe optical-delay line, Prof. Norman Tolk for making the optical parametric amplifier system available for these experiments, and Francis Afzal for the concept drawing in Figure 1(a). Silicon waveguides were fabricated at the Center for Nanophase Materials Sciences, which is a DOE Office of Science User Facility. Selective $VO_2$ deposition and SEM imaging were carried out at the Vanderbilt Institute of Nanoscale Science and Engineering.

**References**

1. Miller, K. J.; Haglund, R. F.; Weiss, S. M., Optical phase change materials in integrated silicon photonic devices: review. *Optical Materials Express* **2018,** *8* (8), 2415-2429.
2. Wuttig, M.; Bhaskaran, H.; Taubner, T., Phase-change materials for non-volatile photonic applications. *Nat Photonics* **2017,** *11* (8), 465-476.
3. Brady, N. F.; Appavoo, K.; Seo, M.; Nag, J.; Prasankumar, R. P.; Haglund, R. F.; Hilton, D. J., Heterogeneous nucleation and growth dynamics in the light-induced phase transition in vanadium dioxide. *Journal of Physics-Condensed Matter* **2016,** *28* (12), 125603.
4. Wegkamp, D.; Herzog, M.; Xian, L.; Gatti, M.; Cudazzo, P.; McGahan, C. L.; Marvel, R. E.; Haglund, R. F., Jr.; Rubio, A.; Wolf, M.; Staehler, J., Instantaneous Band Gap Collapse in Photoexcited Monoclinic $VO_2$ due to Photocarrier Doping. *Physical Review Letters* **2014,** *113* (21), 216401.
5. Tao, Z. S.; Zhou, F. R.; Han, T. R. T.; Torres, D.; Wang, T. Y.; Sepulveda, N.; Chang, K.; Young, M.; Lunt, R. R.; Ruan, C. Y., The nature of photoinduced phase transition and metastable states in vanadium dioxide. *Scientific Reports* **2016,** *6*.
6. Rini, M.; Hao, Z.; Schoenlein, R. W.; Giannetti, C.; Parmigiani, F.; Fourmaux, S.; Kieffer, J. C.; Fujimori, A.; Onoda, M.; Wall, S.; Cavalleri, A., Optical switching in $VO_2$ films by below-gap excitation. *Applied Physics Letters* **2008,** *92* (18), 181904.
7. Briggs, R. M.; Pryce, I. M.; Atwater, H. A., Compact silicon photonic waveguide modulator based on the vanadium dioxide metal-insulator phase transition. *Optics Express* **2010,** *18* (11), 11192-11201.
8. Markov, P.; Marvel, R. E.; Conley, H. J.; Miller, K. J.; Haglund, R. F.; Weiss, S. M., Optically Monitored Electrical Switching in $VO_2$. *Acs Photonics* **2015,** *2* (8), 1175-1182.
9. Ryckman, J. D.; Hallman, K. A.; Marvel, R. E.; Haglund, R. F.; Weiss, S. M., Ultra-compact silicon photonic devices reconfigured by an optically induced semiconductor-to-metal transition. *Optics Express* **2013,** *21* (9), 10753-10763.
10. Joushaghani, A.; Jeong, J.; Paradis, S.; Alain, D.; Aitchison, J. S.; Poon, J. K. S., Wavelength-size hybrid Si-$VO_2$ waveguide electroabsorption optical switches and photodetectors. *Opt Express* **2015,** *23* (3), 3657-3668.
11. Wong, H. M. K.; Yan, Z.; Hallman, K. A.; Marvel, R. E.; Prasankumar, R. P.; Haglund, R. F.; Helmy, A. S., Broadband, Integrated, Micron-Scale, All-Optical $Si_3N_4$/$VO_2$ Modulators with pJ Switching Energy. *ACS Photonics* **2019,** *6* (11), 2734-2740.
12. Cocker, T. L.; Titova, L. V.; Fourmaux, S.; Holloway, G.; Bandulet, H. C.; Brassard, D.; Kieffer, J. C.; El Khakani, M. A.; Hegmann, F. A., Phase diagram of the ultrafast photoinduced insulator-metal transition in vanadium dioxide. *Physical Review B* **2012,** *85* (15), 155120.
13. Miller, K. J.; Hallman, K. A.; Haglund, R. F.; Weiss, S. M., Silicon waveguide optical switch with embedded phase change material. *Optics Express* **2017,** *25* (22), 26527-26536.
14. Marvel, R. E.; Harl, R. R.; Craciun, V.; Rogers, B. R.; Haglund, R. F., Jr., Influence of deposition process and substrate on the phase transition of vanadium dioxide thin films. *Acta Materialia* **2015,** *91*, 217-226.
15. Becker, M. F.; Buckman, A. B.; Walser, R. M.; Lepine, T.; Georges, P.; Brun, A., Femtosecond laser excitation dynamics of the semiconductor-metal phase transition in $VO_2$. *Journal of Applied Physics* **1996,** *79* (5), 2404-2408.
16. Becker, M. F.; Buckman, A. B.; Walser, R. M.; Lepine, T.; Georges, P.; Brun, A., Femtosecond Laser Excitation of the Semiconductor-Metal Phase Transition in $VO_2$. *Applied Physics Letters* **1994,** *65* (12), 1507-1509.
17. Kübler, C.; Ehrke, H.; Huber, R.; Lopez, R.; Halabica, A.; Haglund, R. F.; Leitenstorfer, A., Coherent structural dynamics and electronic correlations during an ultrafast insulator-to-metal phase transition in $VO_2$. *Physical Review Letters* **2007,** *99* (11), 116401.
18. Pashkin, A.; Kübler, C.; Ehrke, H.; Lopez, R.; Halabica, A.; Haglund, R. F.; Huber, R.; Leitenstorfer, A., Ultrafast insulator-metal phase transition in $VO_2$ studied by multiterahertz spectroscopy. *Physical Review B* **2011,** *83* (19), 195120.
19. Morrison, V. R.; Chatelain, R. P.; Tiwari, K. L.; Hendaoui, A.; Bruhacs, A.; Chaker, M.; Siwick, B. J., A photoinduced metal-like phase of monoclinic $VO_2$ revealed by ultrafast electron diffraction. *Science* **2014,** *346* (6208), 445-448.
20. Wall, S.; Wegkamp, D.; Foglia, L.; Appavoo, K.; Nag, J.; Haglund, R. F.; Stahler, J.; Wolf, M., Ultrafast changes in lattice symmetry probed by coherent phonons. *Nature Communications* **2012,** *3*, 721.
21. Wall, S.; Foglia, L.; Wegkamp, D.; Appavoo, K.; Nag, J.; Haglund, R. F.; Stahler, J.; Wolf, M., Tracking the evolution of electronic and structural properties of $VO_2$ during the ultrafast photoinduced insulator-metal transition. *Physical Review B* **2013,** *87* (11), 115126.
22. Tao, Z. S.; Han, T. R. T.; Mahanti, S. D.; Duxbury, P. M.; Yuan, F.; Ruan, C. Y.; Wang, K.; Wu, J. Q., Decoupling of Structural and Electronic Phase Transitions in $VO_2$. *Physical Review Letters* **2012,** *109* (16).
23. Nag, J.; Haglund, R. F.; Payzant, E. A.; More, K. L., Non-congruence of thermally driven structural and electronic transitions in $VO_2$. *Journal of Applied Physics* **2012,** *112* (10), 103532.
24. Laverock, J.; Kittiwatanakul, S.; Zakharov, A. A.; Niu, Y. R.; Chen, B.; Wolf, S. A.; Lu, J. W.; Smith, K. E., Direct Observation of Decoupled Structural and Electronic Transitions and an Ambient Pressure Monocliniclike Metallic Phase of $VO_2$. *Physical Review Letters* **2014,** *113* (21), 216402.
25. Reed, G. T.; Mashanovich, G.; Gardes, F. Y.; Thomson, D. J., Silicon optical modulators. *Nature Photonics* **2010,** *4* (8), 518-526.
26. Lopez, R.; Haglund, R. F.; Feldman, L. C.; Boatner, L. A.; Haynes, T. E., Optical nonlinearities in $VO_2$ nanoparticles and thin films. *Applied Physics Letters* **2004,** *85* (22), 5191-5193.
27. Lin, Q.; Zhang, J.; Piredda, G.; Boyd, R. W.; Fauchet, P. M.; Agrawal, G. P., Dispersion of silicon nonlinearities in the near infrared region. *Applied Physics Letters* **2007,** *91* (2), 021111.
28. Lysenko, S.; Rua, A.; Vikhnin, V.; Fernandez, F.; Liu, H., Insulator-to-metal phase transition and recovery processes in $VO_2$ thin films after femtosecond laser excitation. *Physical Review B* **2007,** *76* (3), 035104.

# ULTRAFAST SWITCHING OF 1550 NM PULSES IN A HYBRID VO$_2$:SILICON WAVEGUIDE

Kent A. Hallman[1], Kevin J. Miller[2], Andrey Baydin[1], Sharon M. Weiss[1,2] and Richard F. Haglund[1]

[1]Department of Physics and Astronomy, Vanderbilt University, Nashville TN 37235-1807 USA

[2]Department of Electrical Engineering and Computer Science, Vanderbilt University, Nashville TN 37235-1807 USA

## SUPPORTING INFORMATION

### 1. CHARACTERIZATION OF THE SIGNAL AND IDLER BEAMS

The spectra of the signal and idler beams are shown in Figure S1(a, c). To obtain these spectra, the optical parametric amplifier (OPA) beam scattered from a beam block placed before the polarizing beam splitter was collected by an optical spectrum analyzer. Note that the optical spectrum analyzer has a wavelength cutoff of 1700 nm, close to the idler wavelength of about 1670 nm. Near this upper wavelength limit, a large signal is seen even in the background spectrum, as seen in the raw spectra in Figure S1. This artifact cannot be removed even by completely covering the optical spectrum analyzer port. To eliminate this artifact, the spectrum in figure S1(b) was obtained by subtracting the two in figure S1(a); subtracting the two signals resulted in the noise seen near 1700 nm in figure S1(b).

Assuming transform-limited pump and probe pulses with a Gaussian temporal distribution, the FWHM of the signal (probe) beam of about 33 nm corresponds to a pulse duration of about 105 fs, while the idler (pump) beam fit of 30 nm yields about 135 fs. Note that the dispersion in the fiber coupling the probe beam to the waveguide stretches the probe pulse to a duration of about 800 fs – a fact that will be significant for the interpretation of the pump-probe experiments.

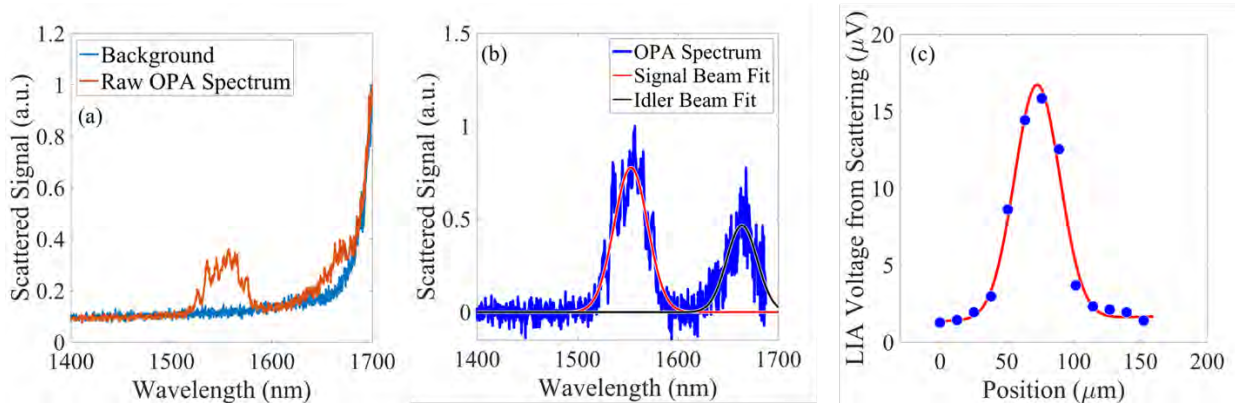

Figure S1: (a) Raw measurement before background subtraction (red) and background spectrum (blue). (b) Pump and probe spectra, extracted from data in (a), with each pulse fit to an appropriate Gaussian function to extract the pulse duration. (c) Spatial profile of pump beam spot.

The beam waist was measured *in situ* using the small amount of pump light scattered into the waveguide. With the probe beam blocked and the lock-in amplifier reference frequency set to 1000

Hz – the pump frequency – the parabolic mirror was translated to sweep the pump beam spot across the device and the signal from the pump light scattered into the waveguide was recorded as a function of beam position. There were negligible differences between measurements in which the parabolic mirror is swept parallel and perpendicular to the waveguide axis; both measurements fit a Gaussian distribution with a $e^{-2}$ beam diameter of ~70 μm. An example of this fit is shown in Figure S1(c). Given the dimensions of the embedded $VO_2$ segment in the waveguide, it is justifiable to treat the pump beam intensity and fluence as constants at the maximum value measured.

## 2. JUSTIFICATION FOR THE PUMP-PROBE CONFIGURATION

An important difference between this out-of-plane pump, in-waveguide probe experiment and a typical pump-probe experiment on thin films is that here the probe beam is modulated to provide the reference frequency to the lock-in amplifier. Conventionally, the pump beam is modulated and optically filtered before the detector so the probe signal is measured at the frequency of the pump. This yields a signal in the lock-in amplifier that corresponds to neither the pump nor the probe separately, but to the *effect of the pump on the sample volume as seen by the probe*. Capitalizing on the fact that phase-sensitive detection can turn small AC voltages into measurable signals, this configuration can routinely measure pump-induced modulations of the sample that produce effects on the differential reflection or transmission of the probe beam as small as $10^{-4}$ to $10^{-6}$.

In this experiment, this level of sensitivity is unnecessary because any device with such a small differential transmission would not be technologically useful. Moreover, unlike the pump-probe geometry for a thin-film measurement, here the pumped and probed volumes are identical. A further advantage of modulating the probe beam, as in Figure 2, is that the signal does not need to be coupled from the lensed fiber into free-beam optics so that an additional optic can be deployed to filter out the pump beam. Although a typical pump-probe configuration for the chopper was used at first, it was found that by chopping the pump beam and filtering it out later, losses associated with the additional optical element (the filter) and coupling back into free-beam reduced the signal-to-noise ratio below that of the configuration where the probe beam is modulated. Instead, the amplitude of the scattered pump light (at 1000 Hz) was sufficiently small that the electronic band-pass filter in the lock-in amplifier (set at 500 Hz) removed it.

An additional, significant advantage of this probe-chopped configuration was ease of alignment. In the probe-chopped configuration, the probe and pump beams could be aligned independently of one another. The pump beam spot position could be aligned to the device by maximizing the pump light scattered into the waveguide and then the coupling of the fiber tips into and out of the waveguides could be optimized by a suitable change of the reference frequency fed to the lock-in amplifier. However, alignment of the more conventional pump-chopped configuration required near-perfect and simultaneous alignment of the fiber-tips and pump beam spot and introduced the additional requirement that the delay stage be near time zero. Given alignment drift on a timescale of a few measurements and the reduced SNR, this would have been very difficult.

However, a few pump-chopped measurements were made to verify that the probe-chopped protocol did not introduce spurious signals, for example, from pump light scattered into the waveguide. These results are shown in Figure S2. Even though the lock-in amplifier time constant was three times as long as for the measurements in the main text (3 s rather than 1 s), inspection of Figure 3(b) still shows better SNR than these pump-chopped measurements. Figure S2 shows a similar large jump in long-time background near 2.6 mJ/cm$^2$ and large jump in amplitude of the transient response around 0.7-0.8 mJ/cm². Due to the poor SNR of the 0.69 mJ/cm² measurement, it is difficult to tell if the *transient* dielectric response is the same. More importantly, the temporal response in the mid-fluence region resembles the probe-chopped measurements in the main text.

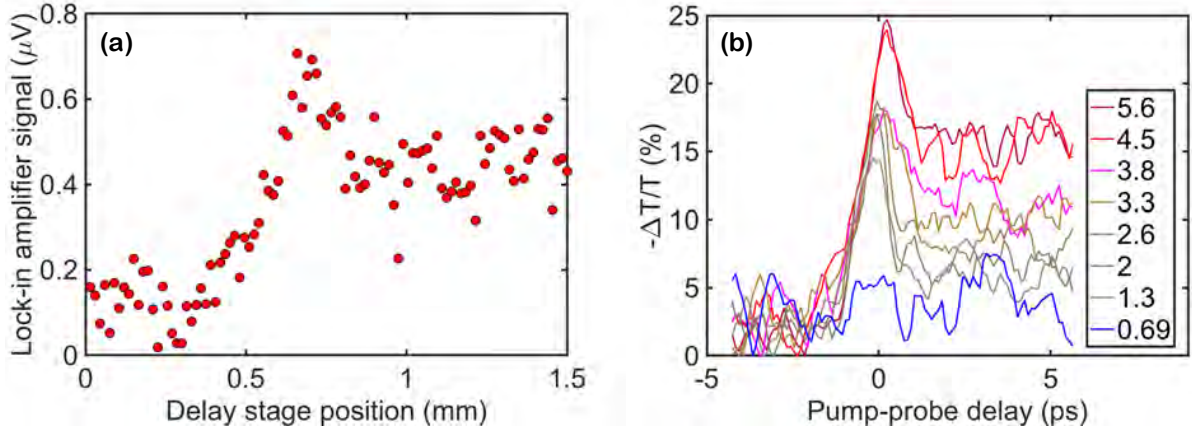

Figure S2: Pump-probe measurement in which the pump is chopped and then optically filtered out before the detector. (a) Example of raw data at 4.5 mJ/cm². (b) Measurements, including the 4.5 mJ/cm² dataset, with position converted to time and waveguide transmission normalized to the unpumped device. The numbers in the inset are the incident fluence in mJ/cm²; color coding is similar to Figure 3(b). The SNR of these measurements is substantially worse than the probe-chopped measurements because of the need to couple the light back into free space in order to utilize the 1550 nm bandpass filter used to isolate the probe beam. The data were smoothed with the same five-point moving average used in Figure 3(b) in the main text.

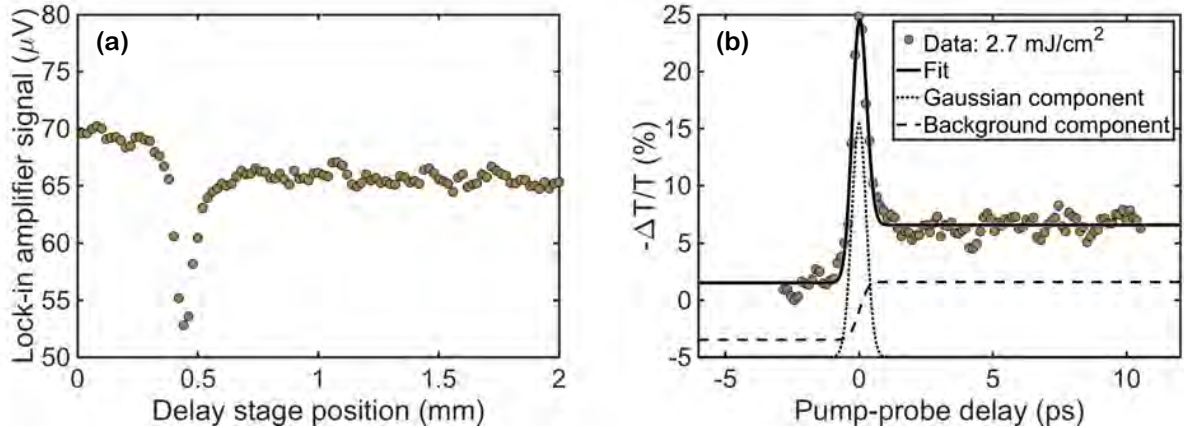

Figure S3: Pump-probe measurement in which the probe is chopped and no optical filtering is used before the detector. (a) Example of raw lock-in amplifier data at 2.7 mJ/cm². (b) The same measurements with position converted to time and waveguide transmission normalized to the unpumped device. The data were not smoothed.

Comparison of the two examples of raw data, Figures S2(a) and S3(a) highlights a final advantage of the probe-chopped measurements. Since the background in the pump-chopped measurement, S2(a), is near zero, drift of the fiber-tips could result in erroneous results. For example, if the fiber tips had become misaligned during the S2(a) measurement at a delay stage position of 0.8 mm, the lock-in amplifier signal would have dropped to zero and it would have appeared that there was no long-time signal. In contrast, if the fiber tips had drifted in the S3(a), it would have become readily apparent because the lock-in amplifier signal would have dropped from 50-70 µV down to about 0 µV. When this, or a more gradual drift, occurred during a probe-chopped measurement, the fiber tips were realigned and the measurement restarted.

The lock-in amplifier signal in Fig. S3(a) also shows that the switched pulse lasts while the delay stage is scanned from approximately 0.25 mm to 0.7 mm. This corresponds to an elapsed transit time for the signal pulse at the LIA of approximately 1.5 ps.

3. **OPTICAL PROPERTIES OF VANADIUM DIOXIDE**

The graphs of refractive index and absorption coefficient in Fig. S4 were calculated from optical constants $n$ and $k$ of a 100 nm $VO_2$ film measured by spectroscopic ellipsometry.

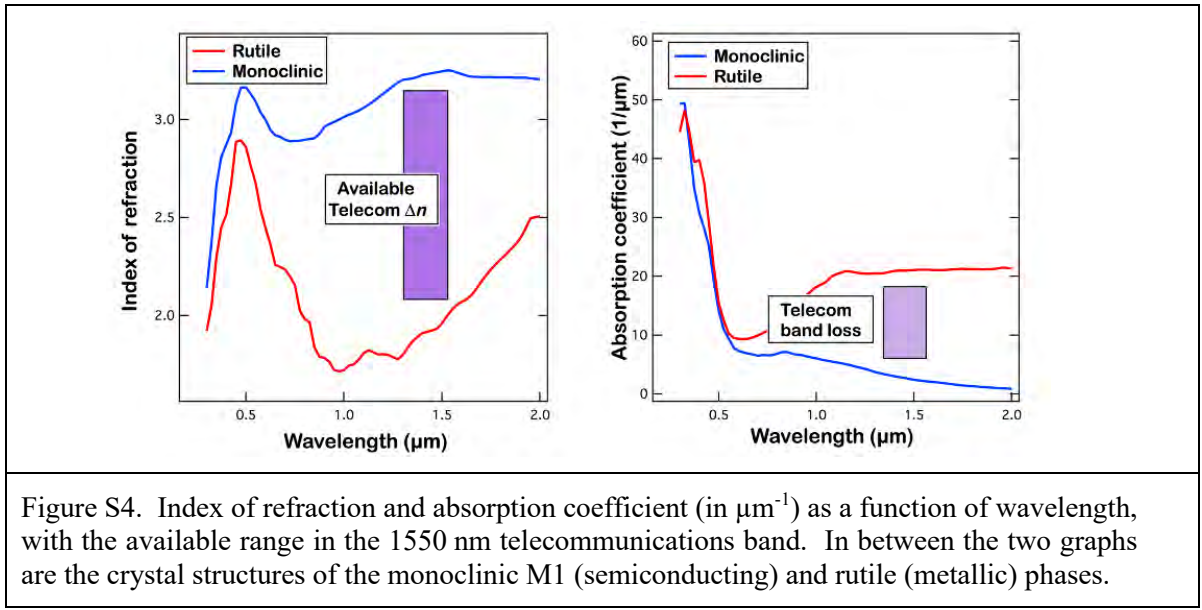

Figure S4. Index of refraction and absorption coefficient (in $\mu m^{-1}$) as a function of wavelength, with the available range in the 1550 nm telecommunications band. In between the two graphs are the crystal structures of the monoclinic M1 (semiconducting) and rutile (metallic) phases.

The nonlinear absorption and refractive index of $VO_2$ have not been measured at 1550 nm, but data for 800 nm, 120 fs pulses are available for 100 nm-thick films shown in Table 2 below.[1]

| | TABLE S1. NONLINEAR ABSORPTION $\beta$ AND REFRACTIVE INDEX $\gamma$ | |
|---|---|---|
| | **MONOCLINIC** | **RUTILE** |
| $\beta$ | 270±30 cm/GW | <1 cm/GW |
| $\gamma$ | -7.1±0.5 cm²/TW | 7.5±0.5 cm²/GW |

4. **CALCULATING ABSORBED FLUENCE AND ENERGY DENSITY**

The absorbed fluence is related to the conventionally quoted incident fluence by calculating the reflection coefficient at the air-$VO_2$ interface:

$$R = \left| \frac{n*-1}{n*+1} \right|^2 = \frac{(n-1)^2 + \kappa^2}{(n+1)^2 + \kappa^2} = \frac{(2.218)^2 + (0.235)^2}{(4.218)^2 + (0.235)^2} \cong \frac{(2.218)^2}{(4.218)^2} = 0.723$$

Assuming negligible scattering at the VO$_2$:Si interface, the ratio of absorbed to incident fluence is approximately

$$\frac{F_{abs}}{F_{inc}} \cong (1-R) = 0.723$$

Thus, an *absorbed* fluence of 1 mJ/cm$^2$ corresponds to the conventional specification of *incident* fluence of 1.38 mJ/cm$^2$.

For the switched pulse shown in Fig. 3(a), the absorbed fluence is very nearly 2 mJ/cm$^2$. The embedded VO$_2$ section in the SOI waveguide structure has a volume (based on an AFM scan) of about 0.150x0.700x0.700 µm$^3$=0.0735 µm$^3$. Assuming that the absorbed fluence is absorbed homogeneously throughout the volume, this would yield an energy cost per switch of 600 fJ for the pump at 1.6 mJ/cm$^2$ and not quite 980 fJ for a pump of 2.7 mJ/cm$^2$.

The volume of the monoclinic VO$_2$ unit cell is 128.4 Å$^3$=128.4·10$^{-12}$ µm$^3$. Hence the embedded volume of VO$_2$ illuminated by the pump laser contains 0.0756/128.4·10$^{-12}$=5.724·10$^8$ unit cells. Given that the absorption coefficient of VO$_2$ at 1670 nm wavelength is of order $\alpha$=2 µm$^{-1}$, the total number of photons absorbed in the embedded VO$_2$ volume of thickness $t$ for an **absorbed** fluence of 1 mJ/cm$^2$ is

$$E_{abs} = \frac{F_{abs}}{t}\left[1-e^{-\alpha t}\right] = \frac{1\text{mJ·cm}^{-2}}{0.150\mu m} \times \frac{10^{-8}cm^2}{\mu m^2}\left[1-e^{-2\cdot 0.15}\right] = 17.3\frac{\text{pJ}}{\mu m^3} = 10.8\cdot 10^7 \frac{\text{eV}}{\mu m^3}$$

The absorbed energy axes on Figures 3(c)-(d) and 4(d) are based on this scaling. The number of photons corresponding to this absorbed energy density is

$$N_{photon} = \frac{10.8\cdot 10^7\text{ eV}/\mu m^3}{0.74\text{ eV}/\text{photon}} \times 0.0735\mu m^3 = 1.07\cdot 10^7 \text{ photons}$$

Assuming that each photon with energy above the band gap, excites an electron into the conduction band, this corresponds to an electron density of order 2.5·10$^{20}$ cm$^{-3}$, near experimentally measured thresholds[2, 3] for exciting the destabilizing 6 THz V-V bond that drives the completion of the monoclinic-to-rutile phase transition. The threshold for completing the structural phase transition, as experimentally measured and calculated in density-functional theory[4], is also of order one photon for several tens of unit cells – consistent with the calculated value for an absorbed fluence of 1 mJ/cm$^2$ in this experiment, which yields a value of 57.24/1.07=~53 unit cells/photon.

5. **EXTRACTING PULSE DURATION DATA FROM MEASUREMENTS IN FIGURE 3(B).**

The relationship between transition rate (as given by Fermi's golden rule #2) and the differential transmission measured in the pump-probe experiment is given by

$$-\Delta\alpha\cdot L = \frac{\Delta T}{T}(\hbar\omega_{probe},t), \Delta\alpha(\hbar\omega_{probe},t) = |M_{fi}|^2 \times \rho_J(\hbar\omega_{probe}) \times \{f_i(t)[1-f_f(t)]\}$$

where $M_{if}$ is the dipole matrix element, $\rho_J$ is the density of states, and $f_i$ ($f_f$) is the fractional population in the initial (final) state of the pumped VO$_2$ segment.

The raw (unsmoothed) data shown in Fig. 3(a), as well as the *unsmoothed* data from which the smoothed data shown in Figure 3(b) were derived, were fit with a Gaussian pulse superimposed on an error function background. That is, only unsmoothed data were fit. The fitting function for the differential transmission is displayed in the following equation:

$$-\frac{\Delta T}{T} = c_1 + c_2 \cdot erf\left(\frac{x-c_3}{\sqrt{2}c_4}\right) + c_5 \cdot e^{-(x-c_3)^2/2c_4^2}$$

The fit function is the sum of a constant and the partial density function (PFD) and cumulative density function (CDF) of a normal distribution. The data sets are not fit globally to allow for independent measurements of the $e^{-2}$ widths, $c_4$, although it is the same for both the PDF and the CDF. In addition, the local fits allow for the slight drift in time-zero seen in Figure 3(b) because the $c_3$ coefficient could vary between measurements.

We define **contrast** as the peak differential transmission less differential transmission at +8 ps where the background is flat except at the highest fluences; the fit function was used to calculate the contrasts plotted in Figure 4(d). This definition emphasizes that the device response of most interest is the response at approximately 1 ps, as discussed in SI Section 2 relating to the duration of the raw probe signal at the lock-in amplifier.

The data were fit using the linear least-squares routine in MatLabR2018. Examples of one fit each from the low-, mid-, and high-fluence regimes are shown in Fig. S5. The data are the same as those in Figure 3(b), but here they are not smoothed since the unsmoothed data were used in the fitting routine.

## 6. COMPILATION OF FITS TO SPECTRA IN LOW-, INTERMEDIATE- AND HIGH-FLUENCE REGIMES.

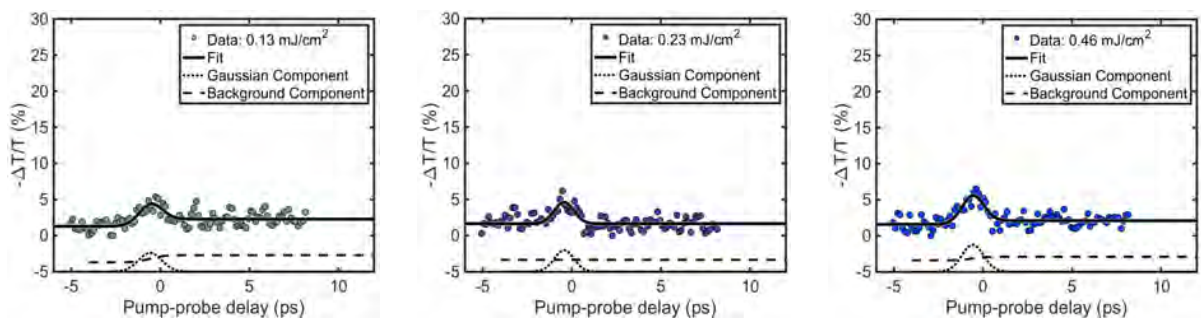

Figure S5. Examples of fits to data for the low-fluence regimes.

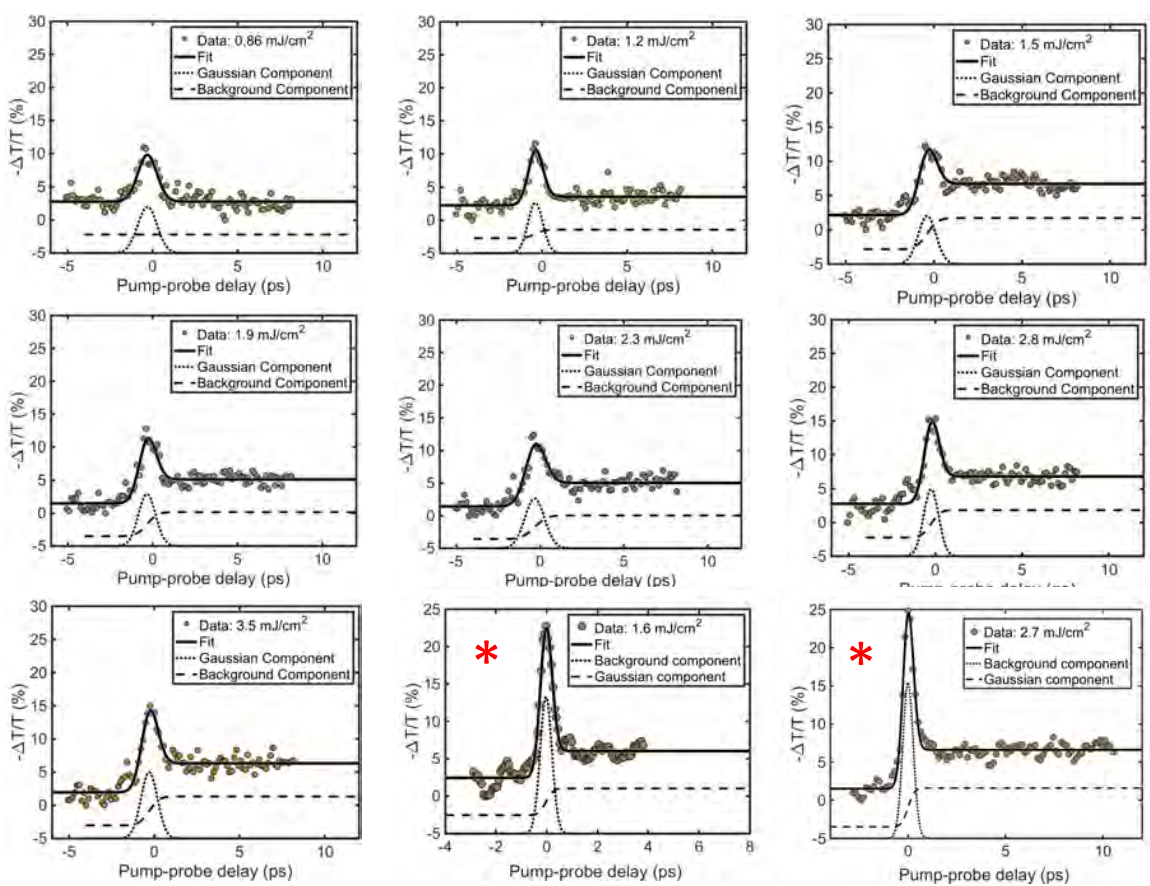

Figure S6. Examples of fits to data for the intermediate-fluence regime. The two spectra marked with an asterisk * were acquired on a different day from the other spectra in this set, with a different alignment. The associated response times were markedly faster, as discussed in connection with the brown diamond data points in Figure 4(d).

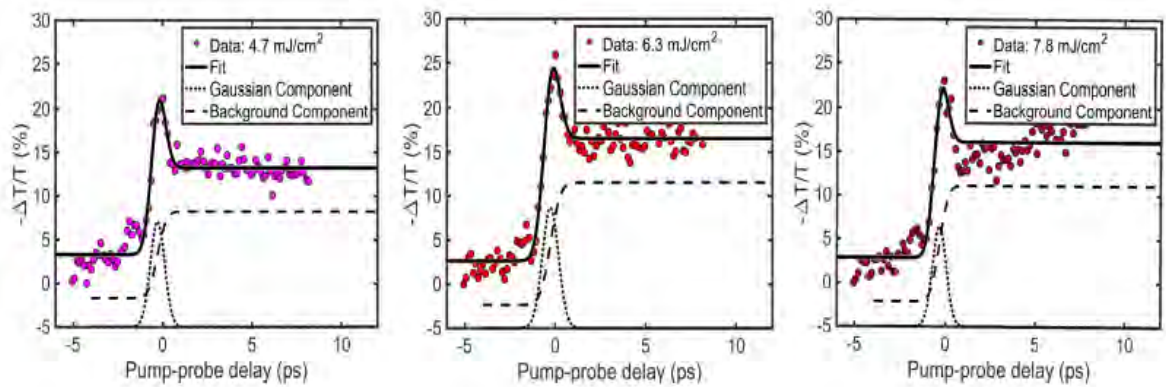

Figure S7. Examples of fits to data for the high-fluence regime.

REFERENCES

1. Lopez, R.; Haglund, R. F.; Feldman, L. C.; Boatner, L. A.; Haynes, T. E., Optical nonlinearities in $VO_2$ nanoparticles and thin films. *Applied Physics Letters* **2004,** *85* (22), 5191-5193.
2. Kübler, C.; Ehrke, H.; Huber, R.; Lopez, R.; Halabica, A.; Haglund, R. F.; Leitenstorfer, A., Coherent structural dynamics and electronic correlations during an ultrafast insulator-to-metal phase transition in $VO_2$. *Physical Review Letters* **2007,** *99* (11), 116401.
3. Wall, S.; Foglia, L.; Wegkamp, D.; Appavoo, K.; Nag, J.; Haglund, R. F.; Stahler, J.; Wolf, M., Tracking the evolution of electronic and structural properties of $VO_2$ during the ultrafast photoinduced insulator-metal transition. *Physical Review B* **2013,** *87* (11), 115126.
4. Appavoo, K.; Wang, B.; Brady, N. F.; Seo, M.; Nag, J.; Prasankumar, R. P.; Hilton, D. J.; Pantelides, S. T.; Haglund, R. F., Jr., Ultrafast Phase Transition via Catastrophic Phonon Collapse Driven by Plasmonic Hot-Electron Injection. *Nano Letters* **2014,** *14*, 1127-1133.